# Maximizing First Order Approximate Mean of SINR under Imperfect Channel State Information for Throughput Enhancement of MIMO Interference Networks

*Ali Dalir [a], Hassan Aghaeinia [b]*

[a] [b] *Department of Electrical Engineering, Amirkabir University of Technology (Tehran Polytechnic), Tehran, Iran.*

*Corresponding Author: alidalir@aut.ac.ir*

## Abstract

In this research paper approximate mean of signal-to-interference-plus-noise ratio (SINR) under imperfect channel state information (CSI) is computed and maximized for throughput enhancement of MIMO interference networks. Each transmitter and receiver has respectively **M** and **N** antennas and network operates in a time division duplex mode. Each transceiver adjusts its filter to maximize the expected value of SINR. The proposed New Approach for Throughput Enhancement under imperfect CSI utilizes the reciprocity of wireless networks to maximize the estimated mean. The sum rate performance of the proposed algorithm is verified using Monte Carlo simulations.

# 1. Introduction

Normally, wireless network scenarios such as interference channel (IC), share the channel among the users, resulting in multi-user interference. Among different medium access control (MAC), such as TDMA, FDMA, CDMA, a new method termed interference alignment (IA), leads to the efficient use of communication resources, since it successfully achieves the theoretical bound on the multiplexing gain. This scheme fits the unwanted signals from other users into a small part of the signal space observed by each receiver (interference subspace) while the other signal subspace is left free of interference for the desirable signal. The alghiment solution has been provided in different papers [1]-[7].

In [3], every transmitter uses a precoder which should be designed to accommodate all the interference signals into one half of the received signal space dimensions and leaves the other half without interference for the desired signal. The long precoder sizes at transmitters show a barrier to implementing such a scheme. Since such assumptions are too hard to materialize, it is very complicated to design a system based on such an elegant scheme. In [8]-[10], the authors propose methods based on the designed scheme of [3] to reduce precoder sizes. Another barrier in [3] is the assumption of global channel knowledge. [11] show that when the direct links have different characteristic functions (channel permutation or memory), in the absence of half part of CSI (cross links), one can achieve full degrees-of-freedom.

In practice, CSI is far from being perfect due to a variety of reasons, such as channel estimation error, quantization error, feedback error / delay, and etc. In [12]-[13] the performance of IA under CSI error was quantified. Mean loss in sum rate compared to perfect CSI case increases unboundedly as SNR increases.

The reliability of IA is little known, which is the subject of [14]. Authors study the error performance of IA. Since most IA algorithms require extensive channel state information (CSI), authors also investigate the impact of CSI imperfection (uncertainty) on the error performance. [14] design bit loading algorithms that significantly improve error performance of the existing IA schemes. Furthermore, [14] propose an adaptive transmission scheme produces robustness to CSI uncertainty to reduce error probability.

*Beamforming Strategy Based on the Interference Alignment:* In order to maximize sum rate of the MIMO interference network, beamforming strategy based on the interference alignment is used. Progressive minimization of the leakage interference is the basis for such algorithms [4, Algorithm 1], [5], and [7]. Other algorithms include Max-SINR algorithm [4, Algorithm 2], and minimum mean square error [6]. These schemes are established based on the availability of perfect CSI. The performance of transceivers is sensitive to CSI inaccuracies. Different algorithms are proposed to improve the throughput of the IC, under imperfect CSI.

*Beamforming Strategy with Imperfect CSI:* Researchers have tried to improve sum rate of the MIMO interference network under imperfect CSI via robust transceiver design. In order to maximize system throughput, beamforming strategy based on the interference alignment is used. In [15], authors applied a minimum mean square error criterion to improve robustness of the MIMO IC for a channel with uncertainty. The authors in [16] proposed a robust distributed joint signal and interference alignment algorithm for the MIMO cognitive radio networks. Interference alignment is evaluated as a technique to mitigate inter-cell interference in the downlink of a cellular network for the case of imperfect channel knowledge [17].

In this research paper approximate mean of signal-to-interference-plus-noise ratio (SINR)

under imperfect channel state information (CSI) is computed and maximized for throughput enhancement of MIMO interference networks. Each transceiver adjusts its filter by maximizing the expected value of SINR.

The contribution of this paper or the presented novelty compared to the previous work in [4] is that approximate mean of SINR over CSI error is used for maximization to enhance throughput of MIMO interference networks. Numerical results demonstrate when approximate mean is used for maximization the proposed transceivers will lead to sum rate improvement.

The convergence of the proposed algorithm is demonstrated. Accuracy of approximation is studied via Monte Carlo simulations. Monte Carlo simulations demonstrate that more accurate approximation can be achieved with less SNR.

## 2. System Model

In a $K$-user MIMO IC, transmitter $j$ and receiver $k$ have $M$ and $N$ antennas, respectively. Independent symbols $D^j$ with power $P$ are sent by the $j^{th}$ transmitter. True and estimated channel matrices between transmitter $j$ and receiver $k$ are denoted by $G^{kj}$ and $H^{kj}$, respectively. Then, the error model is described by:

$$G^{kj} = H^{kj} + E^{kj} . \qquad (1)$$

The elements of $E^{kj}$, error matrix, are independent and identically distributed Gaussian with zero mean and variance $\sigma^2$. The received signal at receiver $k$ is expressed by

$$Y^k = \sum_{j=1}^{K}(H^{kj} + E^{kj})X^j + Z^k , \qquad (2)$$

where $X^j$ is the $M \times 1$ signal vector transmitted by the transmitter $j$ and $Z^k \sim CN(0, N_0 I)$ is additive white Gaussian noise (AWGN) vector. Transmitter $j$ precodes symbol vector by using the precoder matrix. $V^j$ is the $M \times D^j$ precoder matrix. Columns of $V^j$, $v_d^j$, are unit norm vectors. Receiver $k$ estimates the transmitted symbol vector by using the interference suppression matrix $U^k$. The received signal is filtered by $U^k$ as $\overline{Y^k} = U^{k\dagger} Y^k$.

Each node works in a time division duplex (TDD) mode. At two consecutive time slots, first, nodes on the left-hand side send the data to the nodes on the right-hand side. Then the role of nodes is switched and the nodes on the left-hand side receive the data, as illustrated in Fig. 1.

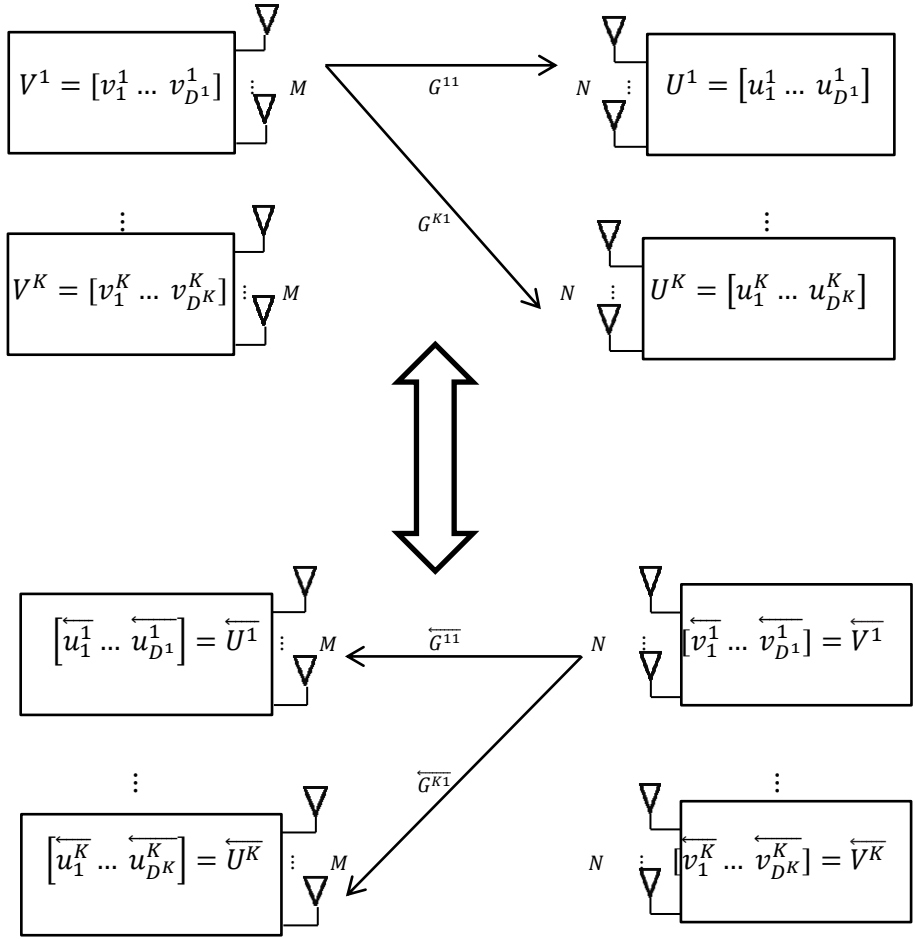

*Fig. 1. System model. Reciprocal network (below channel) is obtained by switching the roles of transmitters and receivers in the original channel (top network). Original and reciprocal channels distinguish two working modes.*

The relation between the original and reciprocal channel matrices is $\overleftarrow{G^{jk}} = G^{kj\dagger}$ [4]. Since the receivers of the reciprocal channel play the roles of original network's transmitters and vice versa, then $\overleftarrow{V^k} = U^k$ and $\overleftarrow{U^j} = V^j$.

## 3. New Approach for Throughput Enhancement under Imperfect CSI

In this section, the proposed algorithm is formulated.

### 3.1. $SINR_d^k$ under Imperfect CSI

According to the system model, the SINR value for the $d^{th}$ data stream at $k^{th}$ receiver is expressed by

$$SINR_d^k = \frac{P\left\|u_d^{k\dagger} G^{kk} v_d^k\right\|^2}{P\sum_{j=1}^{K}\sum_{m=1}^{D^j}\left\|u_d^{k\dagger} G^{kj} v_m^j\right\|^2 - P\left\|u_d^{k\dagger} G^{kk} v_d^k\right\|^2 + N_0 \|u_d^k\|^2}. \quad (3)$$

The random variables, $A$ and $B$, are used to represent signal-to-interference-plus-noise ratio as a rational function $SINR_d^k = \frac{A}{B}$ and are defined by:

$$A = P u_d^{k\dagger}\left[(H^{kk} + E^{kk})v_d^k v_d^{k\dagger}(H^{kk} + E^{kk})^\dagger\right] u_d^k,$$

$$B = u_d^{k\dagger}\left[P \sum_{j=1}^{K}\sum_{m=1}^{D^j}\left((H^{kj} + E^{kj})v_m^j v_m^{j\dagger}(H^{kj} + E^{kj})^\dagger\right)\right.$$
$$\left. - P\left((H^{kk} + E^{kk})v_d^k v_d^{k\dagger}(H^{kk} + E^{kk})^\dagger\right) + N_0 I\right] u_d^k. \quad (4)$$

It is seen that $SINR_d^k$ are functions of error matrices. Next, the mean of $A$, and $B$ are computed as follow

$$\mu_1 = Mean[A] = u_d^{k\dagger}\left[PH^{kk}v_d^k v_d^{k\dagger}H^{kk\dagger} + P\sigma^2 I\right]u_d^k, \tag{5}$$

$$\mu_2 = Mean[B] = u_d^{k\dagger}\left[P\sum_{j=1}^{K}\sum_{m=1}^{D^j} H^{kj}v_m^j v_m^{j\dagger}H^{kj\dagger} - PH^{kk}v_d^k v_d^{k\dagger}H^{kk\dagger} + (P\sigma^2\sum_{j=1}^{K}D^j - P\sigma^2 + N_0)I\right]u_d^k.$$

Where, $Mean[A]$ is obtained as follow. $Mean[B]$ is computed similarly.

$$Mean\left[(H^{kj} + E^{kj})v_m^j v_m^{j\dagger}(H^{kj} + E^{kj})^{\dagger}\right] = Mean\left[H^{kj}v_m^j v_m^{j\dagger}H^{kj\dagger}\right] +$$

$$Mean\left[H^{kj}v_m^j v_m^{j\dagger}E^{kj\dagger}\right] + Mean\left[E^{kj}v_m^j v_m^{j\dagger}H^{kj\dagger}\right] + Mean\left[E^{kj}v_m^j v_m^{j\dagger}E^{kj\dagger}\right] = \tag{6}$$

$$H^{kj}v_m^j v_m^{j\dagger}H^{kj\dagger} + 0 + 0 + \sigma^2\left(v_m^{j\dagger}v_m^j\right)I = H^{kj}v_m^j v_m^{j\dagger}H^{kj\dagger} + \sigma^2 I.$$

In simplistic way, $A$, and $B$ can be approximated by $\mu_1$, and $\mu_2$. Therefore, $SINR_d^k$ with respect to $\mu_1$, and $\mu_2$ is given by

$$SINR_d^k \cong \frac{u_d^{k\dagger}\left[PH^{kk}v_d^k v_d^{k\dagger}H^{kk\dagger} + P\sigma^2 I\right]u_d^k}{u_d^{k\dagger}\left[P\sum_{j=1}^{K}\sum_{m=1}^{D^j}H^{kj}v_m^j v_m^{j\dagger}H^{kj\dagger} - PH^{kk}v_d^k v_d^{k\dagger}H^{kk\dagger} + (P\sigma^2\sum_{j=1}^{K}D^j - P\sigma^2 + N_0)I\right]u_d^k}. \tag{7}$$

*Statistical Linearization Argument*

If $f(A, B)$ is concentrated near its mean, then $E\left[\frac{A}{B}\right]$ can be expressed in terms of $\mu_1 = E[A]$ and $\mu_2 = E[B]$. According to the statistical linearization argument [18], $SINR_d^k$ is approximated by a Taylor series expansion around mean value $(\mu_1, \mu_2)$:

$$SINR_d^k(A, B) \cong SINR_d^k(\mu_1, \mu_2) + \frac{\partial SINR_d^k(\mu_1,\mu_2)}{\partial A}(A - \mu_1) + \frac{\partial SINR_d^k(\mu_1,\mu_2)}{\partial B}(B - \mu_2) +$$

$$\frac{\partial SINR_d^{k^2}(\mu_1,\mu_2)}{\partial A^2}(A - \mu_1)^2 + \frac{\partial SINR_d^{k^2}(\mu_1,\mu_2)}{\partial B^2}(B - \mu_2)^2. \tag{8}$$

In this case:

$$E[SINR_d^k] \cong \frac{\mu_1}{\mu_2} + \int\int \left[\frac{\partial SINR_d^k(\mu_1,\mu_2)}{\partial A}(A-\mu_1) + \frac{\partial SINR_d^k(\mu_1,\mu_2)}{\partial B}(B-\mu_2)\right] f(A,B)dA \times dB$$

$$+ \int\int \left[\frac{\partial SINR_d^{k^2}(\mu_1,\mu_2)}{\partial A^2}(A-\mu_1)^2 + \frac{\partial SINR_d^{k^2}(\mu_1,\mu_2)}{\partial B^2}(B-\mu_2)^2\right] f(A,B)dA \times dB \,. \tag{9}$$

First order estimation of the mean value can be expressed by $E[SINR_d^k] \cong \frac{\mu_1}{\mu_2}$. Second order estimation of the mean value is $E[SINR_d^k] \cong \frac{\mu_1}{\mu_2} + \frac{2\mu_1}{\mu_2^3}VAR[B]$; $\frac{\partial^2}{\partial A^2}\left(\frac{A}{B}\right) = 0$. Therefore, $SINR_d^k$ with respect to $\mu_1$, and $\mu_2$ is first order estimation of the mean. In the context of computation, it is hard to compute $VAR[B]$ theoretically (It is mathematically intractable).

Numerical results demonstrate when first order approximate of the mean is used for maximization the proposed transceivers will lead to sum rate improvement, as shown in Fig. 3, Fig. 4, and Fig. 5. Although first order approximation is maximally within %60 of the true value, proposed scheme achieves higher sum rate compared to baseline schemes considered for comparison.

### 3.2. Algorithm Formulation

The algorithm starts with arbitrary transmit and receive filters and then iteratively updates these filters to provide the solution. The goal is to achieve a robust transceiver by progressively increasing $Mean[SINR]$. The iterative algorithm alternates between the original and reciprocal networks. Within each network, only the receivers update their filters. The algorithm is implemented by following two steps:

*Step I*

*In the original network, the columns of interference suppression filter are updated by each receiver as follow.*

$$\max_{u_d^k} \frac{u_d^{k\dagger}\left[PH^{kk}v_d^k v_d^{k\dagger}H^{kk\dagger}+P\sigma^2 I\right]u_d^k}{u_d^{k\dagger}\left[P\sum_{j=1}^{K}\sum_{m=1}^{D^j}H^{kj}v_m^j v_m^{j\dagger}H^{kj\dagger}-PH^{kk}v_d^k v_d^{k\dagger}H^{kk\dagger}+(P\sigma^2\sum_{j=1}^{K}D^j-P\sigma^2+N_0)I\right]u_d^k}, \quad \forall d \in \{1,\dots,D^k\}.$$

Maximization of (7) over $u_d^k$ can be stated as follow

$$\max \frac{u_d^{k\dagger} Q u_d^k}{u_d^{k\dagger} F u_d^k}, \tag{10}$$

$$Q = Q^\dagger = PH^{kk}v_d^k v_d^{k\dagger}H^{kk\dagger} + P\sigma^2 I \geq 0,$$

$$F = F^\dagger = P\sum_{j=1}^{K}\sum_{m=1}^{D^j}H^{kj}v_m^j v_m^{j\dagger}H^{kj\dagger} - PH^{kk}v_d^k v_d^{k\dagger}H^{kk\dagger} + \left(P\sigma^2\sum_{j=1}^{K}D^j - P\sigma^2 + N_0\right)I > 0.$$

The unit vector that maximizes (7), is given by (Solution is given in [19]. Brief discussion about solution is given next page).

$$u_d^k = \vartheta[F^{-1}Q], \tag{11}$$

operator $\vartheta[.]$ denotes the eigenvector corresponding to the maximal eigenvalue of a matrix.

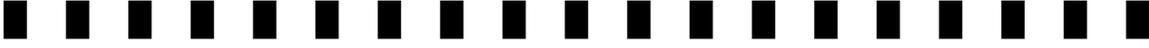

It is shown in [19] that the optimization problem in (10) is equivalent to (12)

$$\max u_d^{k\dagger} Q u_d^k ,$$
$$\text{s. t. } u_d^{k\dagger} F u_d^k = 1 .$$
(12)

For the equivalent problem, i.e. constrained maximization in (12), Lagrangian function can be derived as $l(u_d^k, \lambda) = u_d^{k\dagger} Q u_d^k + \lambda\left(1 - u_d^{k\dagger} F u_d^k\right)$. Lagrange conditions are $\frac{\partial l(u_d^k,\lambda)}{\partial u_d^k} = \mathbf{0}$ and $\frac{\partial l(u_d^k,\lambda)}{\partial \lambda} = 0.$ The solution is denoted by $u_d^{k*}$ and Lagrange multiplier by $\lambda^*$. It is also shown in [19] that $u_d^{k*}$ is the eigenvector corresponding to the maximal eigenvalue of $F^{-1}Q$ and $\lambda^*$ is $u_d^{k*\dagger} Q u_d^{k*}$. To summarize

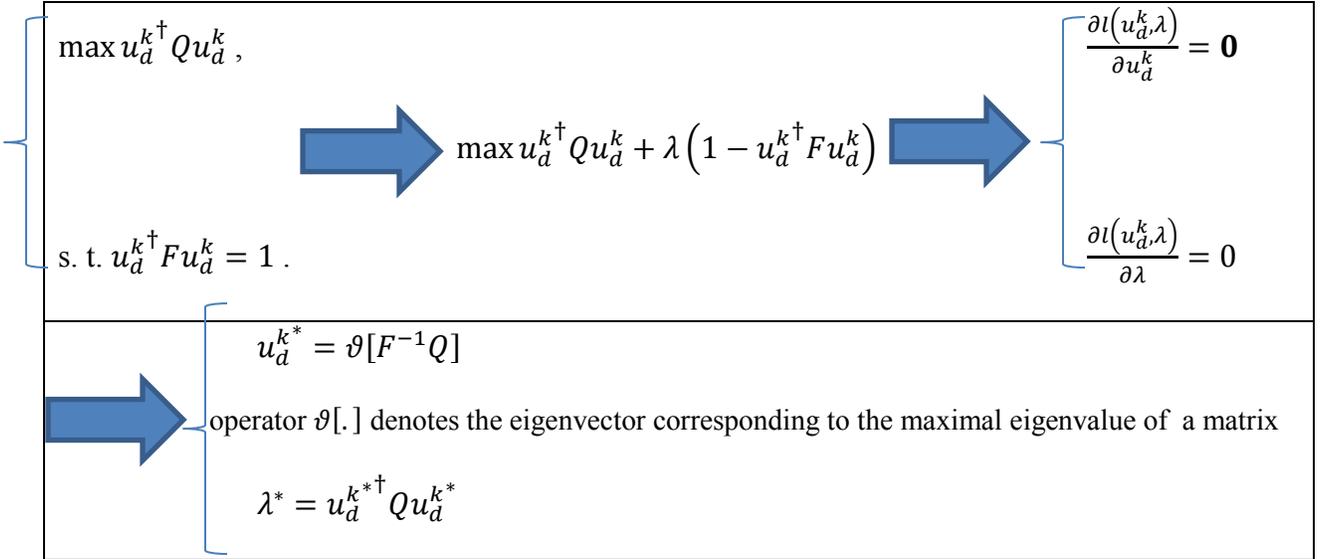

### Step II

*Transmit precoding matrices in the reciprocal network are the receive interference suppression matrices in the original network, determined in Step I. Each receiver solves the*

*following optimization problem:*

$$\max_{\overleftarrow{u_d^J}} Mean[\overleftarrow{SINR_d^J}], \quad \forall d \in \{1, \ldots, D^j\}.$$

The transmit precoding matrices, $\overleftarrow{V^k}$, are the receive interference suppression matrices $U^k$ from the original network that their columns are given by (11). The optimal $d^{th}$ unit column of $\overleftarrow{U^J}$, is given by

$$\overleftarrow{u_d^J} = \vartheta[\overleftarrow{F}^{-1}\overleftarrow{Q}]. \tag{13}$$

Now, receive interference suppression matrices in the reciprocal network, obtained using (13), replace the transmit precoding matrices in the original network, and then the algorithm returns to Step I. The switching between both channels continues in this manner. The steps of the algorithm are given in Fig. 2.

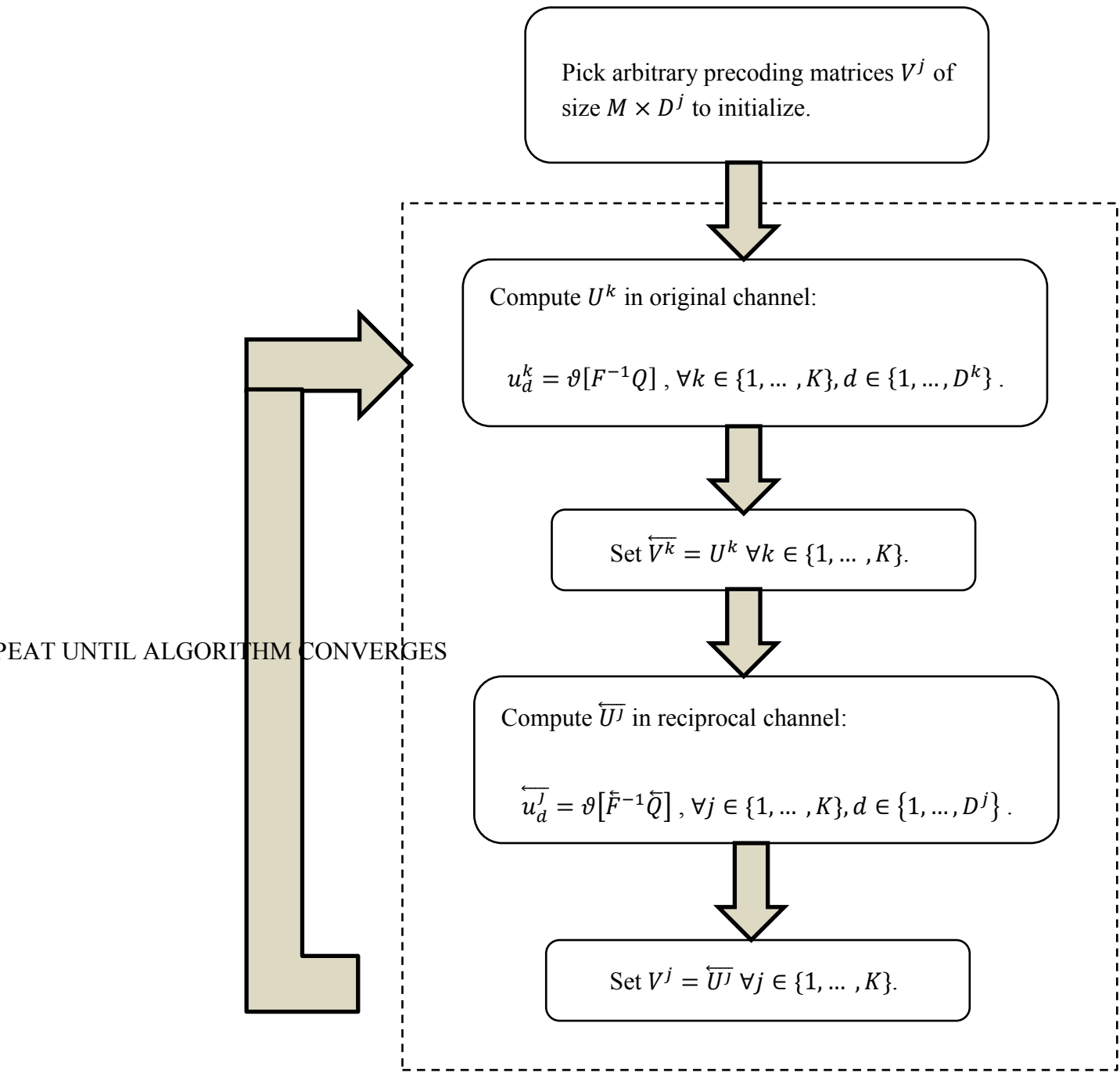

*Fig. 2.* Robust Distributed Transceiver Design.

## 4. Proof of Convergence

Now, the convergence of the proposed algorithm is demonstrated. The metric is defined in (14). It is proved here that each step in the algorithm increases the metric. Since it cannot increase unboundedly, this implies that algorithm converges. It is important to note that the metric is the same for both original and reciprocal networks.

$$\max_{\substack{V^j \text{ and } U^K \\ \forall j \text{ and } k \in \mathcal{K}}} metric = \sum_{k=1}^{K} \sum_{d=1}^{D^k} l(u_d^k, \lambda). \tag{14}$$

Accordingly:

$$\max_{\substack{U^K \\ \forall k \in \mathcal{K}}} metric = \sum_{k=1}^{K} \sum_{d=1}^{D^k} \max_{u_d^k} l(u_d^k, \lambda). \tag{15}$$

In other words, given $V^j \ \forall j \in \mathcal{K}$, Step 1 increases the value of (14) over all possible choices of $U^k \ \forall k \in \mathcal{K}$. The filter $\overleftarrow{U^j}$ computed in Step 3, based on $\overleftarrow{V^k} = U^k$, also maximizes the metric in the reciprocal channel (16).

$$\max_{\substack{\overleftarrow{U^j} \\ \forall j \in \mathcal{K}}} \overleftarrow{metric},$$

$$\overleftarrow{metric} = \sum_{j=1}^{K} \sum_{d=1}^{D^j} \overleftarrow{l}\left(\overleftarrow{u_d^j}, \overleftarrow{\lambda}\right) = \sum_{j=1}^{K} \sum_{d=1}^{D^j} \overleftarrow{u_d^j}^\dagger \overleftarrow{Q} \overleftarrow{u_d^j} + \overleftarrow{\lambda}\left(1 - \overleftarrow{u_d^j}^\dagger \overleftarrow{F} \overleftarrow{u_d^j}\right). \tag{16}$$

Since $\overleftarrow{V^k} = U^k$ and $\overleftarrow{U^j} = V^j$, the metric remains unchanged in the original and reciprocal networks, according to following equation:

$$\overleftarrow{metric} = \sum_{j=1}^{K} \sum_{d=1}^{D^j} u_d^{j\dagger} \left[PH^{jj} v_d^j v_d^{j\dagger} H^{jj\dagger} + P\sigma^2 I\right] u_d^j +$$

$$\sum_{j=1}^{K} \sum_{d=1}^{D^j} \lambda_d^j \left(1 + u_d^{j\dagger} \left[PH^{jj} v_d^j v_d^{j\dagger} H^{jj\dagger} - (P\sigma^2 \sum_{k=1}^{K} D^k - P\sigma^2 + N_0)I\right] u_d^j\right) - \tag{17}$$

$$P \sum_{j=1}^{K} \sum_{d=1}^{D^j} \sum_{k=1}^{K} \sum_{m=1}^{D^k} \lambda_d^j u_m^{k\,\dagger} H^{kj} v_d^j v_d^{j\,\dagger} H^{kj\,\dagger} u_m^k = metric \;.$$

Therefore, Step 3 also can increase the value of (14). Since the value of (14) is monotonically increased after every iteration, convergence of the algorithm is guaranteed.

## 5. Simulation Results

The proposed new approach for throughput enhancement under **i**mperfect CSI is evaluated in this section. Channel coefficients are i.i.d. Gaussian with zero mean and unit variance. We assume quasi-static fading so the fading channels $G^{kj}$ remain unchanged during a fading block. The overall sum rate of the system is given by $R = \sum_{k=1}^{K} \sum_{d=1}^{D^k} R_d^k$ where

*Throughput of $d^{th}$ data stream at $k^{th}$ receiver:* $R_d^k = \log(1 + sinr_d^k)$ ,

$$sinr_d^k = \frac{P \left\| u_d^{k\,\dagger} H^{kk} v_d^k \right\|^2}{P \sum_{j=1}^{K} \sum_{m=1}^{D^j} \left\| u_d^{k\,\dagger} H^{kj} v_m^j \right\|^2 - P \left\| u_d^{k\,\dagger} H^{kk} v_d^k \right\|^2 + N_0 \left\| u_d^k \right\|^2} , \qquad (18)$$

*Overall sum rate:* $= \sum_{k=1}^{K} \sum_{d=1}^{D^k} R_d^k$ .

Fig. 3 represents the sum rate comparison between the proposed and basic algorithms for MIMO IC with $K = 4$ user and $N = M = 3$ antennas and $D = 1$ data stream, $(3 \times 3,1)^4$ MIMO IC. The filters are designed with the error variance of $\sigma^2 = 0.1$. It can be observed that proposed scheme achieves higher sum rate compared to all the other schemes over the entire considered

SNR[1] range. Proposed scheme achieves 7dB SNR gain over the Max-SINR algorithm at providing 14 b/s/Hz sum data rate.

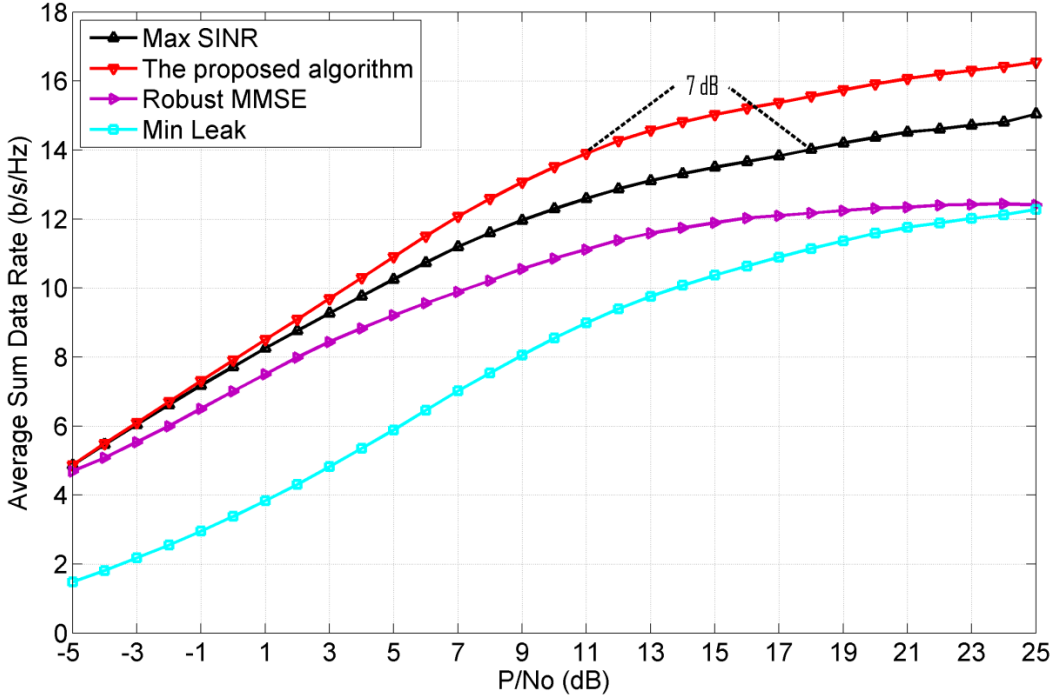

**Fig. 3.** *Average sum data rate versus SNR.* $(3 \times 3, 1)^4$ *MIMO IC* $\sigma^2 = 0.1$.

Fig. 4, and Fig. 5 show sum rate for $(3 \times 4, 2)^2$ (MIMO IC with K=2 user and M = 3 N = 4 antennas and D = 2 data stream) and $(2 \times 2, 1)^3$ (MIMO IC with K=3 user and N = M = 2 antennas and D = 1 data stream). Again, proposed scheme achieves higher sum rate compared to all the other schemes. In comparison with Max-SINR, proposed scheme improves data rate better than 16 b/s/Hz, while Max-SINR cannot achieve data rate higher than 12 b/s/Hz in $(3 \times 4, 2)^2$ MIMO IC. Comparative sum rate improvement compared to Max-SINR is shown for $(2 \times 2, 1)^3$ MIMO IC.

---

1. $\frac{P}{N_0}$ is SNR in the network, since all data streams are of power $P$ and $N_0$ is noise power at all receivers.

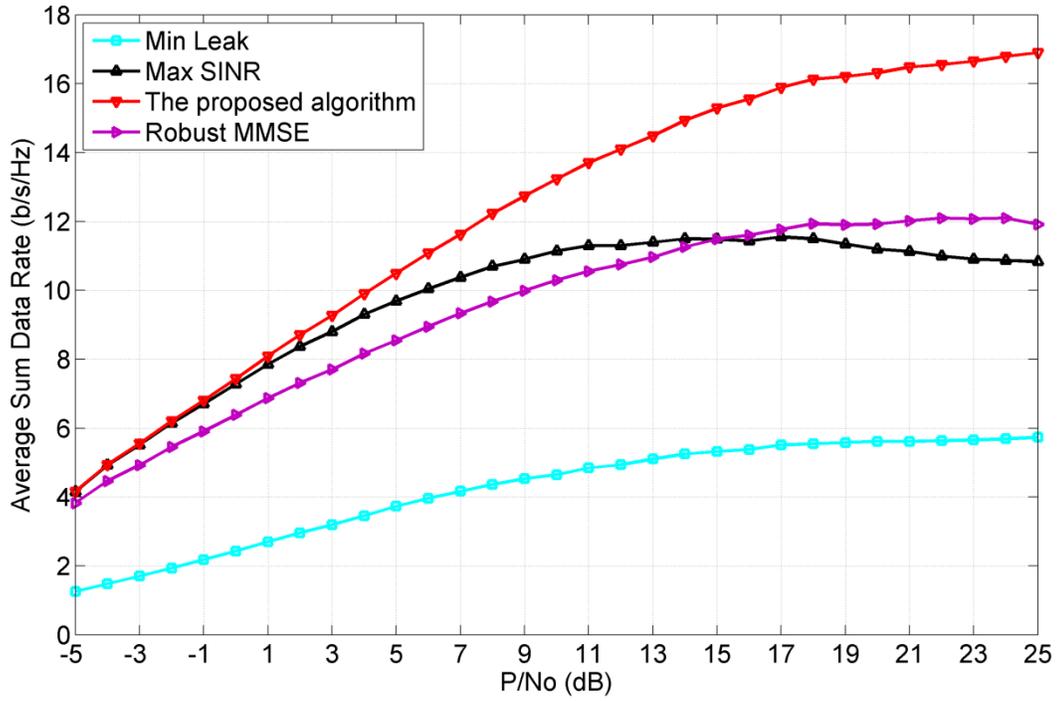

*Fig. 4. Average sum data rate versus SNR.* $(3 \times 4, 2)^2$ *MIMO IC* $\sigma^2 = 0.1$.

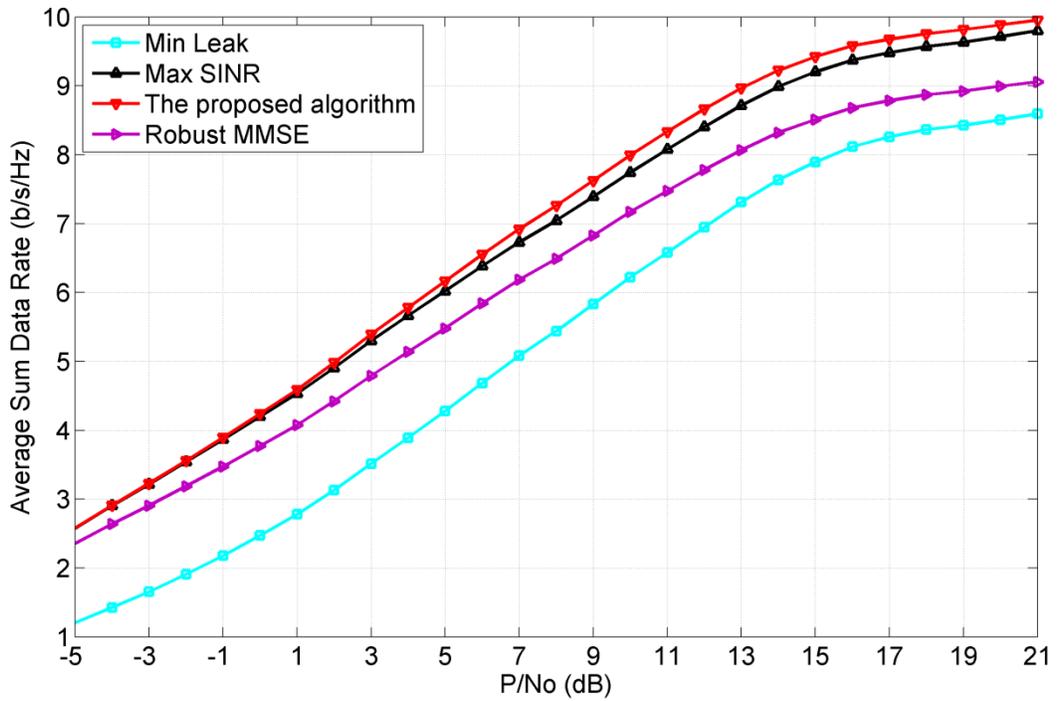

*Fig. 5. Average sum data rate versus SNR.* $(2 \times 2, 1)^3$ *MIMO IC* $\sigma^2 = 0.1$.

Fig. 6, Fig. 7, and Fig. 8 represent the Energy Efficiency (=Sum rate/consumed power) comparison between the proposed and basic algorithms. Figures 6, 7, 8 report higher Energy Efficiency for the proposed algorithm compared to the other schemes.

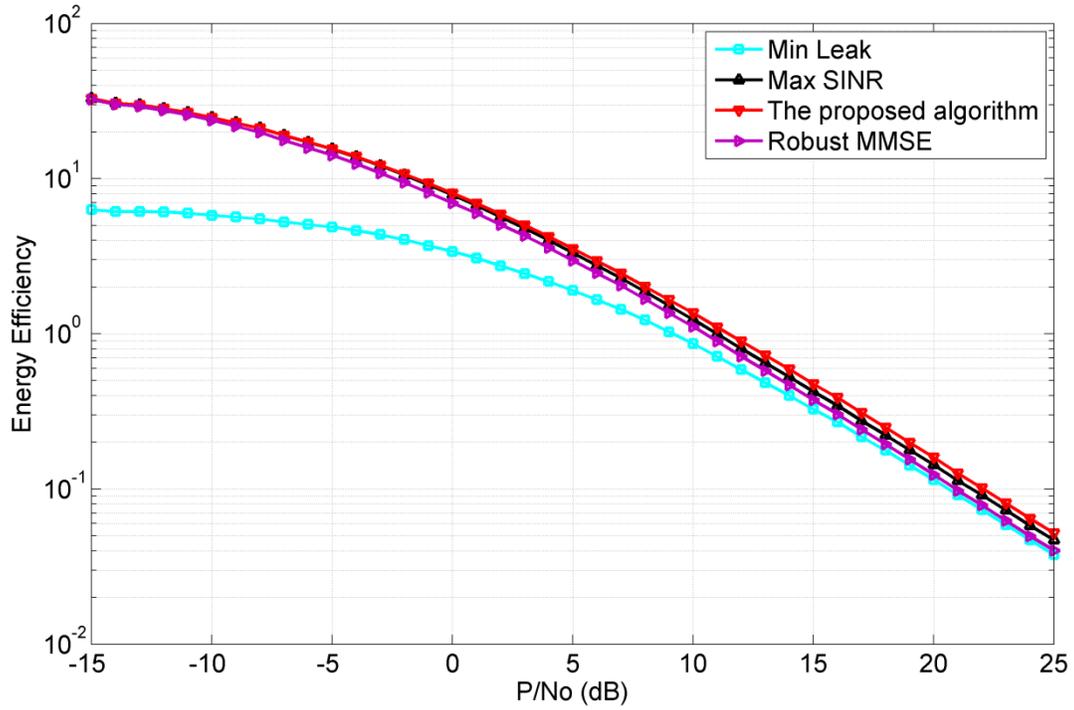

***Fig. 6.*** *Energy Efficiency versus SNR.* $(\mathbf{3 \times 3, 1})^{\mathbf{4}}$ *MIMO IC.*

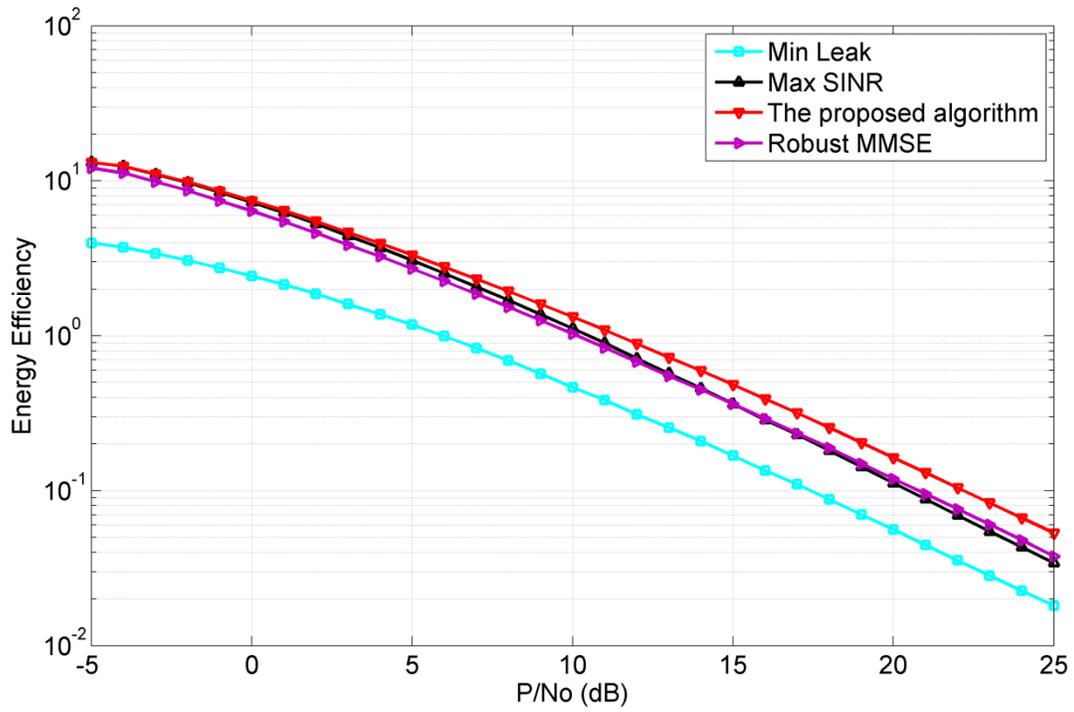

*Fig. 7. Energy Efficiency versus SNR.* $(\mathbf{3 \times 4, 2})^{\mathbf{2}}$ *MIMO IC.*

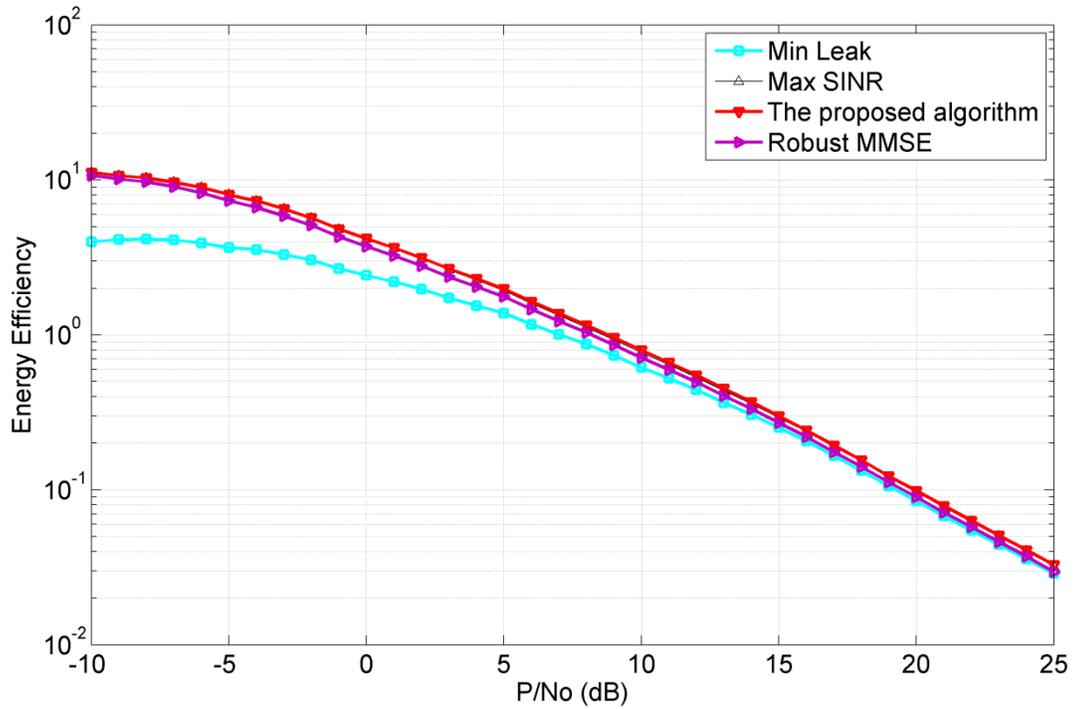

*Fig. 8. Energy Efficiency versus SNR.* $(\mathbf{2 \times 2, 1})^{\mathbf{3}}$ *MIMO IC.*

Numerical value of $Mean[SINR_d^k]$ and theoretical approximation are depicted for $(3 \times 3,1)^4$ MIMO IC in Fig. 9, $(3 \times 4,2)^2$ in Fig. 10, and $(2 \times 2,1)^3$ in Fig. 11. The network uses proposed algorithm to compute precoding and interference suppression matrices. The filters are designed with error variances $\sigma_1^2 = 0.05$ and $\sigma_2^2 = 0.1$.

SNR scales linearly with $\sigma^2$ as it is obvious from (7). It is straightforward to say as $P\sigma^2$ decreases, the impact of any error in approximating diminishes. Here the impacts of influential parameters on the accuracy of approximation are confirmed via Monte Carlo simulations. Accuracy of approximation is measured by $\alpha = \frac{\text{Num} - \text{App}}{\text{Num}}$. Table I, Table II, and Table III shows any approximation error will be attenuated as SNR decreases.

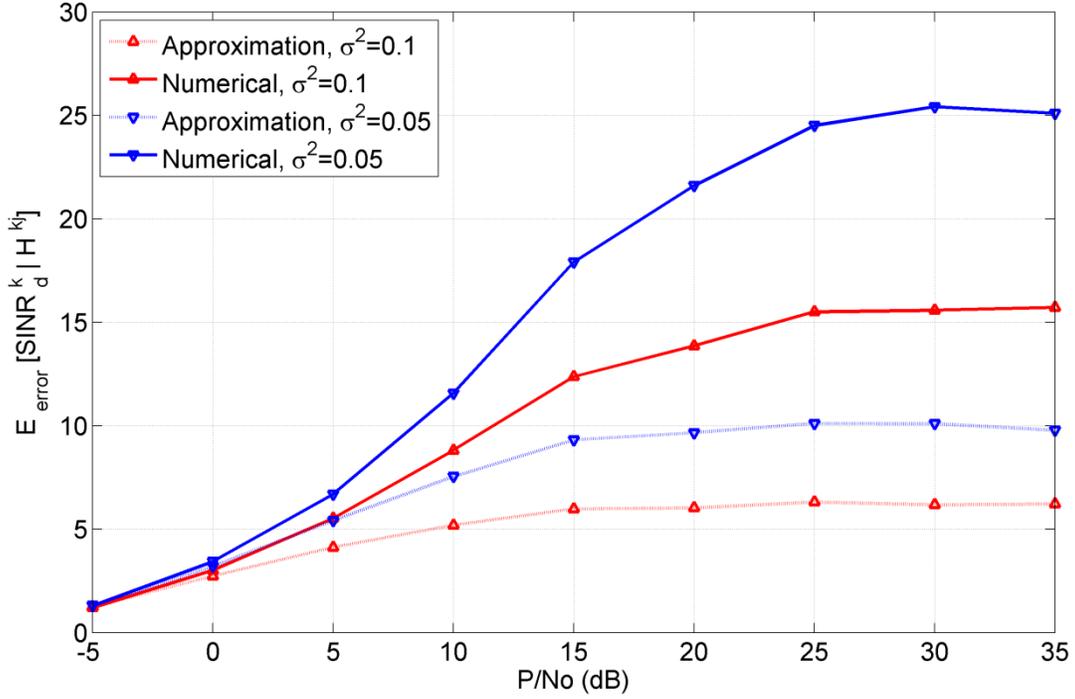

*Fig. 9.* *Approximation of* $Mean[SINR_d^k]$ *and numerical value is shown versus SNR.* $(3 \times 3, 1)^4$ *MIMO IC. The filters are designed with proposed algorithm and two CSI error variances* $\sigma_1^2 = 0.05$ *and* $\sigma_2^2 = 0.1$.

**Table I**

*Accuracy of approximation in Fig. 6.*

| PdB | | | | | | | | | | |
|---|---|---|---|---|---|---|---|---|---|---|
| -5 | 0 | 5 | 10 | 15 | 20 | 25 | 30 | 35 | $\sigma^2$ | |
| 3.18 | 9.49 | 25.22 | 41.1 | 51.64 | 56.5 | 59.28 | 60.37 | 60.49 | = 0.1 | $\frac{Num-App}{Num} \times 100$ |
| 1.53 | 6.61 | 19.03 | 34.78 | 47.97 | 55.26 | 58.75 | 60.28 | 61.02 | = 0.05 | |

*Although accuracy of approximation is higher than %50 ($\alpha > 0.5$) for SNR > 15 dB, it was observed in Fig. 3 that proposed scheme achieved higher sum rate compared to all the other schemes over the entire SNR range.*

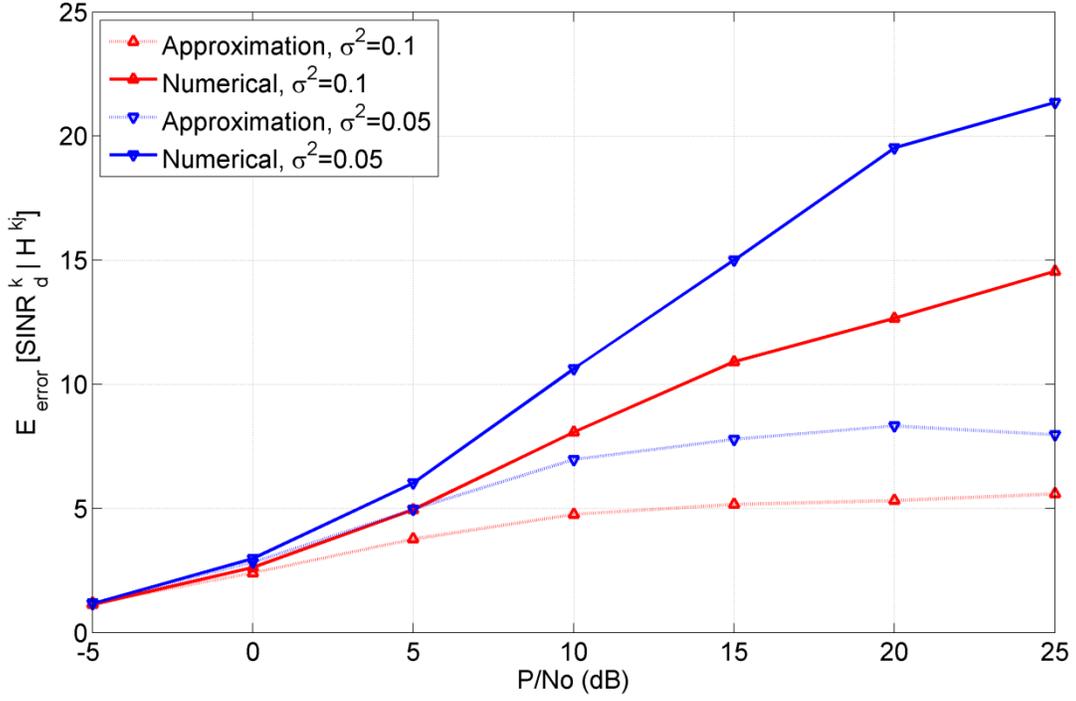

***Fig. 10.*** *Approximation of* $Mean[SINR_d^k]$ *and numerical value is shown versus SNR.* $(3 \times 4, 2)^2$ *MIMO IC.*

**Table II**

*Accuracy of approximation in Fig. 7.*

| | | | PdB | | | | | |
|---|---|---|---|---|---|---|---|---|
| -5 | 0 | 5 | 10 | 15 | 20 | 25 | $\sigma^2$ | |
| 4.46 | 8.37 | 23.76 | 41 | 52.65 | 57.96 | 61.61 | $= 0.1$ | $\dfrac{Num - App}{Num} \times 100$ |
| 2.07 | 4.97 | 17.53 | 34.44 | 48.16 | 57.37 | 62.69 | $= 0.05$ | |

*Although accuracy of approximation extends to %63 in the considered SNR range of Table II, it was observed in Fig. 4, that proposed scheme achieved higher sum rate compared to all the other schemes.*

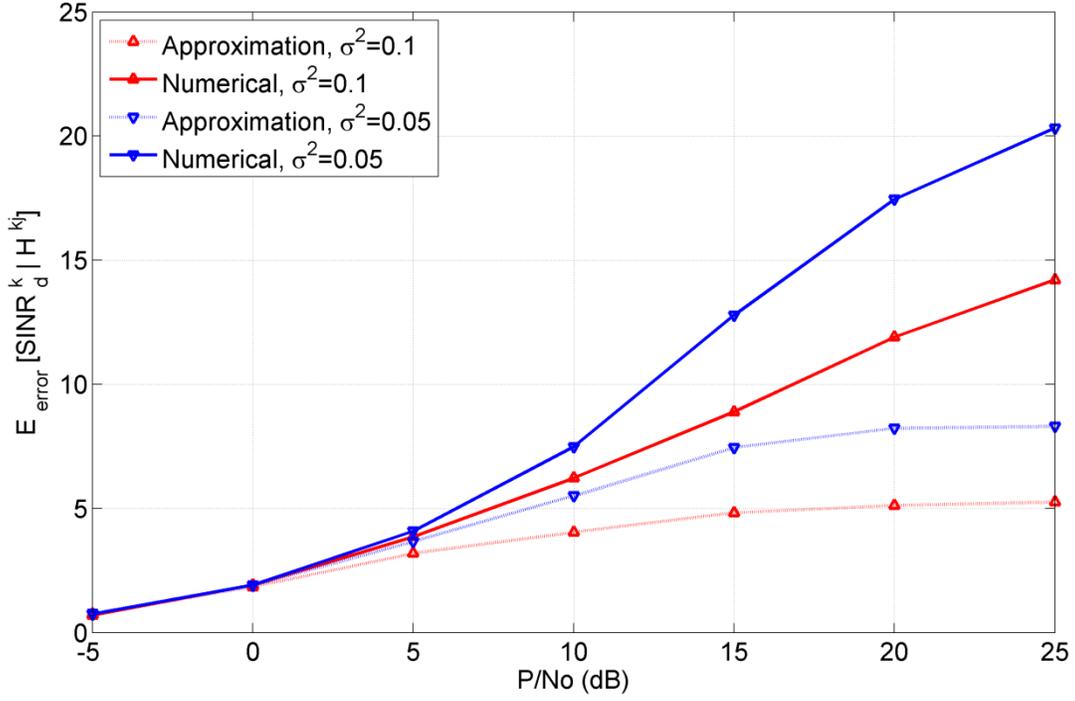

***Fig. 11.*** *Approximation of* $Mean[SINR_d^k]$ *and numerical value is shown versus SNR.* $(2\times 2,1)^3$ *MIMO IC.*

**Table III**

*Accuracy of approximation in Fig. 8.*

| PdB | | | | | | | | |
|---|---|---|---|---|---|---|---|---|
| -5 | 0 | 5 | 10 | 15 | 20 | 25 | $\sigma^2$ | |
| 7.25 | 3.93 | 17.4 | 34.95 | 45.72 | 56.95 | 63.02 | $=0.1$ | $\frac{Num-App}{Num}\times 100$ |
| 3.54 | 1.36 | 10.32 | 26.44 | 41.71 | 52.84 | 59.13 | $=0.05$ | |

*Although accuracy of approximation extends to %63 in the considered SNR range of Table III, it was observed in Fig. 5, that proposed scheme achieved higher sum rate compared to all the other schemes.*

## 6. Conclusion

In this paper, a robust algorithm was proposed to improve the throughput of the MIMO interference channel, under imperfect CSI. The effect of CSI imperfection on the SINR mean was approximated. In the proposed new approach for throughput enhancement of MIMO interference networks under imperfect CSI, filters were adjusted based on the problem of SINR expectation maximization. Transceivers were designed based on the reciprocity of wireless networks. Monte Carlo simulations demonstrated that the proposed algorithm improves data rate of MIMO IC under imperfect CSI.